\documentstyle[psfig]{mn2e}

\newif\ifAMStwofonts

\input{psfig.sty}
\def\mb{\mbox{}}
\def\be{\begin{equation}}
\def\ee{\end{equation}}
\title{The minimum period problem in cataclysmic variables}
\author[J. Barker, U. Kolb]{J.~Barker and U.~Kolb \\Department of Physics \& Astronomy, The Open University, Walton Hall, Milton Keynes, MK7~6AA}

\date{Submitted to MNRAS: July 2002}

\begin{document}

\maketitle

\begin{abstract}
We investigate if consequential angular momentum losses (CAML) or an intrinsic deformation of the donor star in CVs could increase the CV bounce period from the canonical theoretical value $\sim65$ min to the observed value $P_{min} \approx77$ min, and if a variation of these ef\mb fects in a CV population could wash out the theoretically predicted accumulation of systems near the minimum period (the period spike). We are able to construct suitably mixed CV model populations that a statisticial test cannot rule out as the parent population of the observed CV sample. However, the goodness of f\mb it is never convincing, and always slightly worse than for a simple, f\mb lat period distribution. Generally, the goodness of f\mb it is much improved if all CVs are assumed to form at long orbital periods. The weighting suggested by King, Schenker \& Hameury (2002) does not constitute an improvment if a realistically shaped input period distribution is used.
\end{abstract}


\section{Introduction}
\label{introduction}

Cataclysmic variables (CVs) are short-period binaries containing a white dwarf (WD) primary (with mass $M_1$) and a low mass main sequence secondary (with mass $M_2$). The secondary f\mb ills its Roche lobe and transfers mass to the WD through the inner Lagrangian ($L_1$) point.

\begin{figure}
  \psfig{file=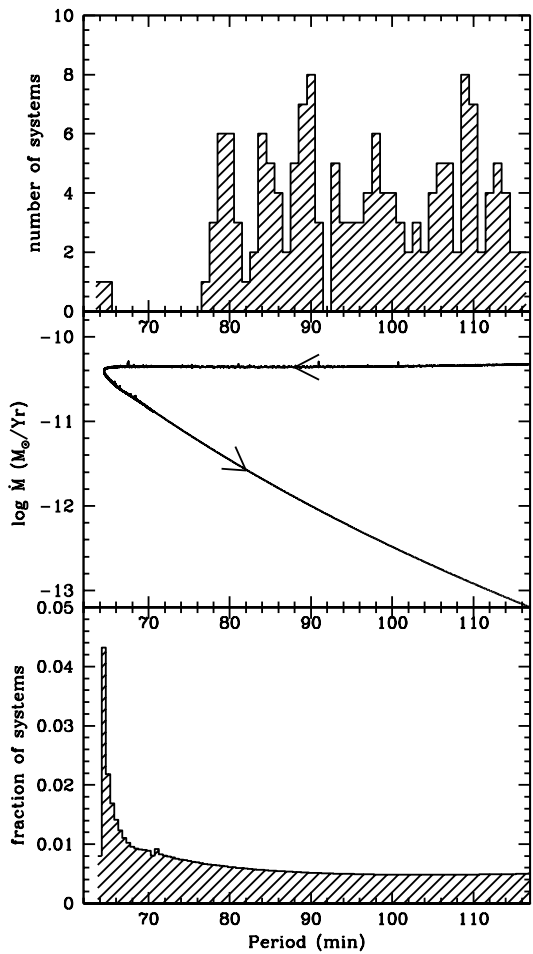,width=\columnwidth}
  \vspace{-0.8cm}
  \caption{ 
    Upper frame: The observed period distribution of CVs with periods less than 116 minutes. Middle frame: Calculated evolutionary track in the orbital period versus mass transfer rate ($\dot{M}$) plane. Lower frame: Period distribution expected from evolutionary track in middle frame.  
    }
  \label{combined}
\end{figure}

The main features of the orbital period distribution of CVs with hydrogen rich donors are the lack of systems in the 2-3 hr period range (the so-called period gap) and the sharp cut of\mb f of the distribution at around 77 minutes, as can be seen in Figure~\ref{combined} (upper frame; e.g. Ritter \& Kolb 1998).

So far theoretical models have been unable to reproduce the precise position of the observed short-period cut-of\mb f and observed shape of the CV orbital period distribution near this cut-of\mb f. This is summarised in Figure \ref{combined}. Systems that evolve under the inf\mb luence of gravitational radiation (GR; Kraft et al. 1962) as the only sink of orbital angular momentum (AM) reach a minimum period at $P_{min} \sim 65$ minutes (Figure\ref{combined}, middle frame; Paczy\'nski 1971; Kolb \& Baraf\mb fe 1999).The probability of f\mb inding a system within a given period range is proportional to the time taken to evolve through this region. We thus have 
\be
	N(P)\propto\frac{1}{\dot{P}},
\ee
for the number $N(P)$ of systems found within a given orbital period range around $P$, and $\dot{P}$ is the secular period derivative at this period. We thus expect an accumulation of systems (a spike) at $P_{min}$ where $\dot P = 0$ (Figure \ref{combined}, lower frame), while no such spike is present in the observed distribution (Figure\ref{combined}, upper frame).

The orbital period evolution ref\mb lects the radius evolution of the mass donor, which in turn is governed by two competing ef\mb fects. Mass transfer perturbs thermal equilibrium and expands the star. Thermal relaxation reestablishes thermal equilibrium and contracts the star back to its equilibrium radius. The minimum period occurs where the two corresponding time scales, the mass transfer time $t_M$ and the thermal (Kelvin-Helmholtz) time $t_{KH}$ are about equal (e.g.\ Paczy\'nski 1971; King 1988). If $t_M \gg t_{KH}$ then the star is able to contract in response to mass loss, but if $t_M \ll t_{KH}$ the star will not shrink rapidly enough and will become oversized for its mass. The position of the minimum period is therefore af\mb fected by the assumed mass transfer rate, and in particular by the assumed rate of orbital angular momentum (AM) losses. In this paper we investigate ways to increase the period minimum by increasing the mass transfer rate, and investigate ways to ``hide'' the spike by introducing a spread of $P_{min}$ values in the CV population. In particular, we study the ef\mb fect of a form of consequential AM loss (CAML) where the AM is lost as a consequence of the mass transferred from the secondary, i.e. $\dot{J}_{CAML}\propto\dot{M}_2$ (see e.g.\ Webbink 1985).

In section \ref{theory} we outline our general model assumptions and introduce the prescription for CAML. In section \ref{sec22} we present detailed calculations of the long-term evolution of CVs, and in section \ref{comptest} we compare the observed short period CV period distribution with various theoretically synthesized model distributions based on the calculations in section 2.

\section{Theoretical versus observed minimum period}

In this section we investigate possible solutions to the mismatch between the theoretical and observed minimum orbital period in CVs.  

\subsection{CAML description}
\label{theory}

The orbital AM loss rate $\dot{J}$ of a CV can be written as the sum of two terms, 

\be
  \dot{J}=\dot{J}_{sys}+\dot{J}_{CAML},
\ee

where $\dot{J}_{sys}$ denotes the ``systemic'' AM loss rate, such as gravitational wave radiation, that is independent of mass transfer, while $\dot{J}_{CAML}$ is an explicit function of the mass transfer rate. We have  

\be
  \frac{\partial \dot{J}_{sys}}{\partial \dot{M}_2}=0
\ee

and

\be
  \dot{J}_{CAML}\to0\qquad\mbox{as}\qquad\dot{M}_2\to0
\ee

We consider the general case in which the CAML mechanism, along with nova mass ejections, causes a fraction of the transferred mass to leave the system. This fraction may be greater than unity as the primary may lose more mass during a nova outburst than was accreted since the last outburst. 

We employ a generic prescription of the ef\mb fect of a CAML mechanism, thus avoiding the need to specify its physical nature. Possible CAML mechanisms include a magnetic propeller, i.e.\ a system containing a rapidly spinning magnetic WD where some of the transferred material gains angular momentum from the WD spin by interaction with the WD's magnetic f\mb ield (see e.g.\ Wynn, King \& Horne 1997), and an accretion disc wind (see e.g. Livio \& Pringle 1994).
 
Our CAML prescription largely follows the notation of King \& Kolb (1995). The AM is assumed to be lost via mass loss that is axis-symmetrical with respect to an axis A f\mb ixed at the WD centre but perpendicular to the orbital  plane.

We def\mb ine $\alpha$ as the total fraction of mass lost from the secondary that leaves the system. We assume further that a fraction $\beta$ ($0\le\beta\le\alpha$) of the transferred mass leaves the system with some fraction $f$ of the angular momentum it had on leaving the $L_1$ point. 

We also consider mass that is lost from the system via nova mass ejections, which over the long term can be considered as an isotropic wind from the primary (see e.g.\ Kolb et al.\ 2001). This material will carry away the specif\mb ic orbital angular momentum of the primary and will account for the  fraction ($\alpha-\beta$) of the mass loss. We thus obtain

\be
  \dot{J}_{CAML}=\eta b^2\omega\dot{M}_2+\frac{\alpha M_2J\dot{M}_2}{M_1M},
\ee

where we def\mb ine $\eta=\beta f$ as the CAML ef\mb f\mb iciency. For comparison with King \& Kolb (1995) we equate this to  

\be\label{eq:jdotcaml}
 \dot{J}_{CAML}=\nu\frac{\dot{M}_2}{M_2}J,\qquad\nu~>~0,
\ee
and obtain

\be\label{eq:nufinal}
  \nu=\eta(1+q)\left(\frac{b}{a}\right)^2+\frac{\alpha q^2}{(1+q)}.
\ee

For our calculations shown below we use the approximation

\be
  \frac{b}{a}\approx1-\varphi+\frac{\varphi^2}{3}-\frac{5\varphi^4}{12},\quad\mbox{with}\quad \varphi^3=\frac{q}{3+3q}.   
\ee

This is an adaptation of the expression given in Kopal (1959) and is accurate to within 1\% over the range $0<q\le0.4$. 

\subsection{Results of numerical experiments}\label{sec22}
In this subsection we present calculations of the long-term evolution of CVs as they approach and evolve beyond the period minimum. For the computations we used the stellar code by Mazzitelli (1989), adapted to CVs by Kolb \& Ritter (1992). Some of these evolutionary sequences are the basis for the theoretical CV period distributions we present in section~\ref{comptest} below.  

\subsubsection{Consequential angular momentum loss}
\label{minpnumeric}
\begin{figure}
\psfig{file=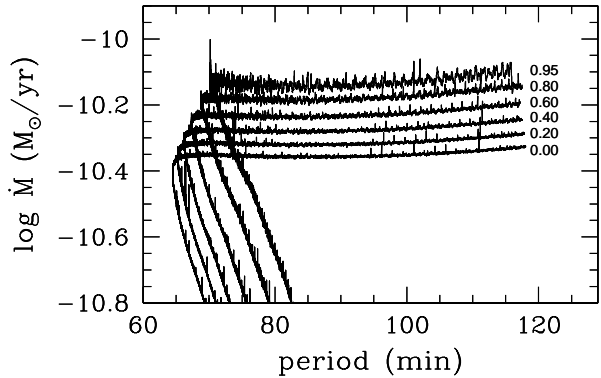,width=\columnwidth}
\caption{
The increase in mass transfer rate and corresponding increase in minimum period for increasing CAML ef\mb f\mb iciency (from 0 to 0.95 as indicated). 
}
\label{fig:fullcaml}
\end{figure}

We calculated the evolution of individual systems that are subject to CAML according to equations \ref{eq:jdotcaml} and \ref{eq:nufinal}. We chose $M_1=0.6M_\odot$ and initial donor mass $M_2=0.2M_\odot$, with a range of CAML ef\mb f\mb iciencies $0\le\eta\le0.95$ as shown in Figure~\ref{fig:fullcaml}.

The systems initially evolve from longer periods towards the period bounce (right to left) at almost constant mass transfer rate. The minimum period increases with increasing CAML ef\mb f\mb iciency to a maximum of around 70 min for $\eta=0.95$.

Mass transfer stability sets an upper limit on the CAML ef\mb f\mb iciency. An obvious upper limit is 1, where all the angular momentum of the transferred material is ejected from the system. Although the ejected material may carry more angular momentum than was transferred (as in the case of a propeller system where additional angular momentum is taken from the spin of the WD) this does not af\mb fect the net loss of orbital angular momentum. 

The maximum CAML ef\mb f\mb iciency still compatible with mass transfer stability could be smaller than unity. The stability parameter $D$ which enters the expression for steady-state mass transfer, equation \ref{eq:stab} (e.g. King \& Kolb 1995) must be greater than zero; this def\mb ines an upper limit on $\eta$.

\be \label{eq:stab}
   -\dot{M}_2=M_2\left(\frac{\dot{J}/J}{D}\right)
\ee

\begin{figure}
\psfig{file=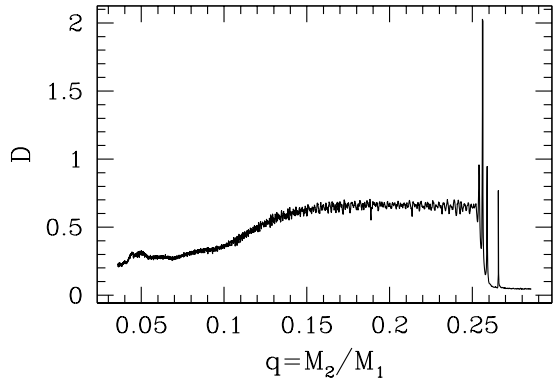,width=\columnwidth}
\caption{
Evolution of the stability factor $D$ with mass ratio $q$ for an initially marginally stable system, ($M_1=0.7 M_\odot$, $M_2=0.2 M_\odot$, $\eta=1.0$). 
}
\label{fig:dq}
\end{figure}

\begin{figure}
\psfig{file=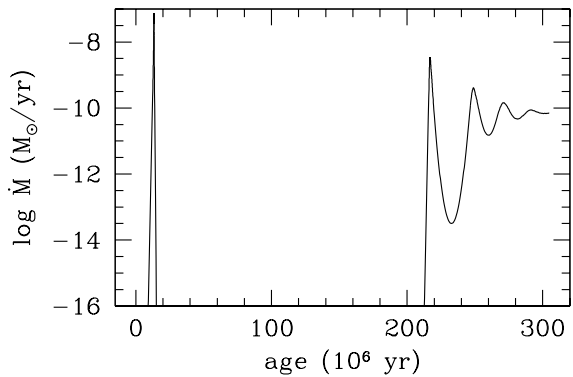,width=\columnwidth}
\caption{Mass transfer rate cycles for the initially marginally stable 
system as in Figure \ref{fig:dq} 
}
\label{fig:highmdot}
\end{figure}

A plot of $D$ against $q$ for an initially marginally stable system ($M_1=0.7M_\odot$, $M_2(init)=0.2M_\odot$ and $\eta=1.0$) is given in Figure \ref{fig:dq}. The system initially exhibits cycles of high mass transfer rate $\dot{M}_2>10^{-9}M_\odot/yr$ ($D$ close to 0) and very low mass transfer rate $\dot{M}_2\to0$. The high states are short lived, on the order of $2\times10^6$ years (see Figure \ref{fig:highmdot}). The system f\mb inally stabilizes with $D\approx0.65$. At around $P_{min}\mbox{ }(q\simeq0.15)$ $D$ starts to decrease further but always remains positive, settling at a value around $0.3$.

\subsubsection{Structure of the secondary}
\label{baraffe1}
The tidal deformation of the secondary may have an ef\mb fect on the period minimum. Calculations by Renvoiz\'e, Baraf\mb fe, Kolb \& Ritter (2002), [see also Kolb 2002] using 3--dimensional SPH models suggest that the secondary is deformed in the non-spherical Roche lobe such that its volume--equivalent radius is around 1.06 times that of the same star in isolation.

We mimic this ef\mb fect in our 1-dimensional stellar structure code by multiplying the calculated radius by a deformation factor $\lambda$ before the mass transfer rate is determined from the dif\mb ference between the radius and the Roche lobe radius via 

\be
  -\dot{M}_2=\dot{M}_0\exp\left(-\frac{R_L-R_2}{H_p}\right).
\ee

Here $\dot{M}_0\simeq10^{-8}M_\odot yr^{-1}$ is the mass transfer rate of a binary in which the secondary just f\mb ills its Roche potential and $H_p$ is the photospheric pressure scale height of the secondary (see e.g.\ Ritter 1988).

\begin{figure}
\psfig{file=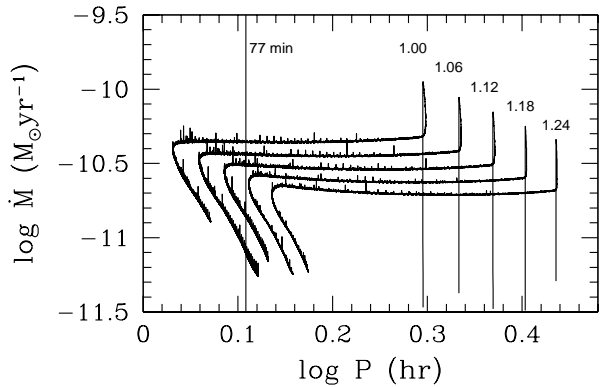,width=\columnwidth}
\caption{
Evolutionary track of a system with $M_1=0.6M_\odot$ and $M_2(init)=0.2M_\odot$ for various deformation factors (as indicated).
}
\label{fig:barraffe}
\end{figure}

Figure~\ref{fig:barraffe} shows the ef\mb fect on the minimum period and mass transfer rate for systems with various deformation factors $\lambda$, ranging from 1 (no deformation) to 1.24. The mass transfer rate is seen to decrease with increasing deformation. This can be understood from the functional dependence  on orbital period and donor mass in the usual quadrupole formula for the AM loss rate due to gravitational radiation (see e.g.\ Landau \& Lifschitz 1958). Although the quadrupole formula is strictly valid only if both
components are point masses, Rezzolla, Ury\=u \& Yoshida (2001) found that the GR rate obtained using a full 3-dimensional representation of the donor star dif\mb fers from the point--mass approximation by less than a few percent.

It can be seen from the f\mb igure that with the deformation factor 1.06 the minimum period increases from around 65 min to around 69 min, consistent with Renvoiz\'e et al (2002) for geometrical ef\mb fects alone. A deformation factor of around 1.18 was required to raise the minimum period to the observed value of $\sim77$ min. This is somewhat larger than the intuitive expectation 
\be
\mbox{increase in radius}\approx\left(\frac{\mbox{new period}}{\mbox{old period}}\right)^{\frac{2}{3}}=\left(\frac{77}{65}\right)^{\frac{2}{3}}=1.12 
\ee
from Kepler's law and Roche geometry.

In our calculations we consider the simple case in which only the geometrical deformation ef\mb  fects are taken into account. The inclusion of the thermal ef\mb fects considered by Renvoiz\'e et al (2002) have the likely ef\mb fect of reducing $P_{min}$, possibly by around 2\% compared to the case with purely geometrical ef\mb fects

One possible physical mechanism that could cause a deformation factor above the value of 1.06 is magnetic pressure inside the star, as suggested by D'Antona (2000). 

We note that Patterson (2000) claims to f\mb ind observational evidence for ``bloated'' secondaries in short period CVs. On the basis of donor mass estimates from the observed superhump excess period he f\mb inds that the donors have $15-30\%$ larger radii than predicted from 1 dimensional., non--deformed stellar models if gravitational radiation is the only AM sink. Even if true, this observation cannot distinguish between an intrinsic deformation of the donor star or the non-equilibrium caused by orbital AM losses in excess of the GR rate.  

\section{Parent distributions versus observations}
\label{comptest}
To test the statistical signif\mb icance of the theoretically predicted accumulation of systems near the period minimum (``period spike'') we calculated the period distributions of model populations for various assumptions about evolutionary parameters. For each parameter a series of evolutionary tracks were generated, typically around 20. 

As systems evolve after the minimum period a point is reached (typically when $\dot{M}$ falls below $10^{-11} (M_{\odot}yr^{-1})$) where numerical f\mb luctuations in $\dot{M}$ become so large that the Henyey scheme no longer converges. The stellar code uses tables to interpolate/extrapolate the opacities and equation of state for each iteration, and in this region the extrapolations become very uncertain. To extend the tracks we used a semi-analytical method as follows.

The tracks were terminated at a value of $\log\dot{M}_2=\log\dot{M}_2(\mbox{Pturn})-0.3$, where $\dot{M}_2(\mbox{Pturn})$ is the mass transfer rate at the minimum period for the track. The radius of the star for the f\mb inal part of the track is approximated by 

\be
   R_2=R_0M_2^{\zeta},
\ee

where $R_0$ and $\zeta$ are assumed to be constant. The values of $R_0$ and $\zeta$ were determined from the f\mb inal few data points for each track. ($\zeta$ takes a typical value of around 0.15 for systems beyond the period bounce.) To generate the extension to the track we then calculated $P$ from the Roche lobe condition, and $\dot{M}_2$ by assuming stationarity as in section \ref{minpnumeric} (see Figure \ref{combined}, middle frame for an example of an extended track).

We weight the chances of observation to the brighter systems by assuming

\be
  \frac{\dot{M}^{\gamma}}{\dot{P}},\qquad\gamma\ge1.0.
\ee

for the detection probability. We tested the calculated model parent distributions for various values of the free parameter $\gamma$ against the observed CV period distribution. A K-S (Kolmogorov-Smirnov) test is insensitive to the dif\mb ferences between the parent distributions. The greatest dif\mb ference in the cumulative distribution functions (CDFs) of the observed and modelled distributions occur at the boundaries of the CDFs, i.e.\ in the least sensitive region for the K-S test (Press et al 1992). We thus decided to use the following modif\mb ied $\chi^2$ test.

\subsection{The modif\mb ied $\chi^2$ test}
\label{chitest}

For each parent distribution 10000 model samples each containing 134 systems were generated. \footnote{134 was the number of observed CVs in the range $77 \le P(min) \le 116$; Ritter \& Kolb 1998, internal update June 2001, as of July 2002 the number of systems in this period range is now 152 though this does alter the values given by the $\chi^2$ test, the trends and hence the results remain unaltered} Each sample was tested against the model parent distribution using a $\chi^2$ test, with 1, 2 and 4 minute bins. This range bridges the need for good resolution and signif\mb icance of the $\chi^2$ test which requires a minimum number of CVs per bin. The observed period distribution was tested against the model parent distribution also, giving the reduced $\chi^2$ value $\chi^2_{obs}$. The fraction $f$ of generated samples with a reduced $\chi^2$ value less than $\chi^2_{obs}$ was used as a measure of the signif\mb icance level of rejecting the hypothesis that the observed distribution is drawn from the parent distribution. In the following we quote the rejection probability Pr=$f$. 

\subsection{Magnetic and non-magnetic CVs}
\label{magnonmag}
\begin{figure}
  \psfig{file=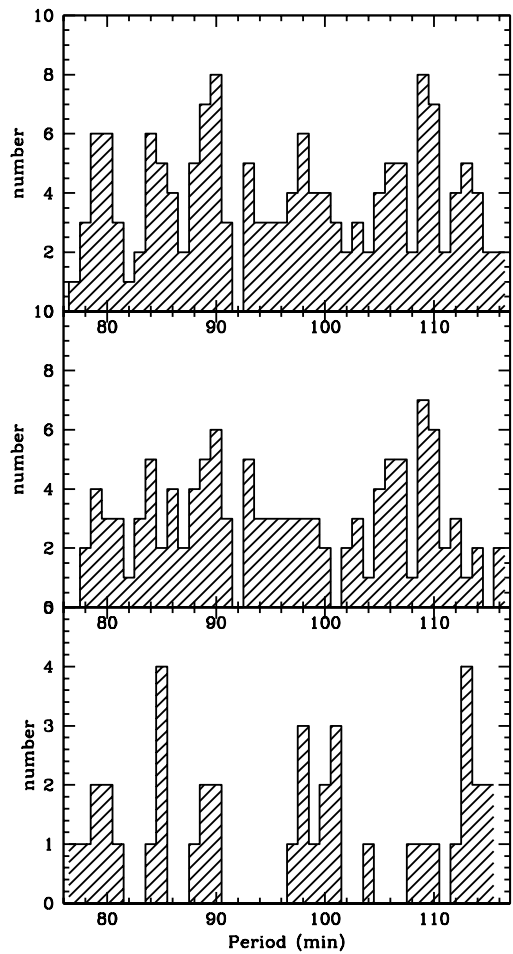,width=\columnwidth}
  \caption{
	Observed orbital period distribution $76 \le P(min) \le  116$. Upper frame: all CVs; Middle frame: non-magnetic CVs; Lower frame: magnetic CVs. 
}
\label{fig:mnmcvs}
\end{figure}

Kolb \& Baraf\mb fe (1999) noted that the observed distribution of non-magnetic CVs (Figure \ref{fig:mnmcvs}, middle frame), and the observed distribution of magnetic CVs (Figure \ref{fig:mnmcvs}, lower frame) show no signif\mb icant dif\mb ference below the period gap. To test and quantify this we compared these distributions for $P\le116$ min, giving a reduced $\chi^2$ probability of 0.1213. Hence we cannot rule out that the distributions are drawn from the same underlying parent distribution. This is borne out by the results of
comparing both distributions with a parent distribution that is f\mb lat in $P$ (see also Table \ref{tab:tabcomb}, entries F and G) which give similar rejection probabilities (Pr=0.709 and Pr=0.781, respectively). We thus f\mb ind no signif\mb icant dif\mb ference between the two distributions. In the following we therefore test models against the combined magnetic and non-magnetic distribution of observed systems.

The lack of any distinct features in the combined observed period distribution (Figure \ref{fig:mnmcvs}, upper frame) does indeed suggest an essentially f\mb lat distribution for the underlying parent distribution. The f\mb lat distribution gives Pr = 0.552 (for the 1 minute bin width, see Table \ref{tab:tabcomb}). We use this value as a benchmark for the models discussed below. 

\begin{table*}
\begin{minipage}{\textwidth}
\begin{center}
\caption{$\chi^2$ tests on various parent distributions.}
\label{tab:tabcomb}
\begin{tabular}{ccccccc}
\hline
 &\multicolumn{2}{c}{1 minute binning}&\multicolumn{2}{c}{2 minute binning}&\multicolumn{2}{c}{4 minute binning}\\
Distribution & $\chi^2_{obs}$ & rejection probability & $\chi^2_{obs}$
 & rejection probability & $\chi^2_{obs}$ & rejection probability\\ 
\hline
A & 1.003 & 0.552 & 0.728 & 0.207 & 0.738 & 0.324\\
B & 1.077 & 0.659 & 0.840 & 0.332 & 0.744 & 0.329\\
C1 & 1.253 & 0.831 & 1.252 & 0.754 & 1.889 & 0.947\\
C3 & 1.065 & 0.609 & 0.887 & 0.380 & 1.148 & 0.669\\
D1 & 1.379 & 0.916 & 1.422 & 0.896 & 1.991 & 0.961\\
D3 & 1.122 & 0.691 & 0.913 & 0.441 & 0.938 & 0.525\\
E1 & 1.211 & 0.786 & 1.258 & 0.779 & 1.503 & 0.848\\
E3 & 1.052 & 0.580 & 0.942 & 0.454 & 0.807 & 0.361\\
F & 1.107 & 0.709 & 1.339 & 0.854 & 1.514 & 0.876\\
G & 1.155 & 0.781 & 1.089 & 0.636 & 1.268 & 0.753\\
\hline
\end{tabular}
\end{center}
\end{minipage}
\end{table*}

\begin{table*}
\begin{center}
\begin{minipage}{\textwidth}
\begin{center}
\begin{tabular}{ll}
\hline
\multicolumn{2}{c}{ KEY}\\
\hline
 A: Flat distribution versus total observed. & D3: M1 versus total observed, $\gamma=3$.\\
 B: Age limit versus total observed. & E1: CAML plus M1 versus total observed, $\gamma=1$.\\
 C1: CAML versus total observed, $\gamma=1$. & E3: CAML plus M1 versus total observed, $\gamma=3$.\\
 C3: CAML versus total observed, $\gamma=3$. & F: Flat distribution versus magnetic CVs only.\\
 D1: M1 versus total observed, $\gamma=1$. & G: Flat distribution versus non-magnetic CVs only.\\
\hline
\end{tabular}
\end{center}
\end{minipage}
\end{center}
\end{table*}

\subsection{Parent populations}
\label{parentpop}
We def\mb ine a standard set of assumptions for simple parent population models as follows:\\ 
(1) The primary mass in all systems is $0.6M_\odot$. This is the value around which the majority of WDs in CVs are expected to form (see e.g.\ de Kool 1992).\\ 
(2) All systems form as CVs at orbital periods greater than 2 hours. This is consistent with the secondary stars in CVs being somewhat evolved (see Baraf\mb fe \& Kolb 2000).\\ 
(3) The f\mb lux of systems through the period gap is constant. That is, suf\mb f\mb icient time has elapsed since the formation of the Galaxy for a steady state to have been reached, so that the number of systems arriving at the lower edge of the period gap is just balanced by the number of new systems forming at orbital periods greater than two hours.\\ 
(4) The CAML ef\mb f\mb iciency is set to 0\\
(5) The systemic AM loss rate is $\dot{J}_{sys}=3\dot{J}_{GR}$, and the deformation factor is $\lambda=1.06$, so that $P_{min}$ equals the observed $P_{min}=77min$, thus enabling us to test the statistical signif\mb icance of the spike.\\ 
(6) Brightness selection factor $\gamma=1.0$\\
A model population subject to these standard assumptions can be rejected with the probability $P_r > 1 - 10^{-4}$.

In the following discussion of various population models we just quote any dif\mb ferences of individual models from this standard set of assumptions.  

\subsubsection{Age limit hypothesis}
\label{agelim}
\begin{figure}
  \psfig{file=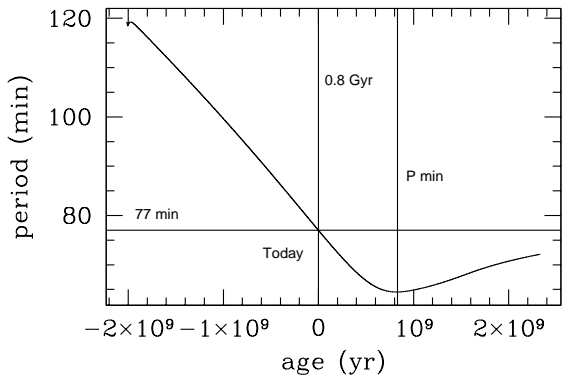,width=\columnwidth}
  \caption{ 
    The age limit hypothesis. A system with current orbital period of 77 minutes will continue to evolve to the true period bounce at around 66 minutes for the next $8\times10^8$ years. 
    }
  \label{fig:agelim}
\end{figure}

It has been suggested that the currently observed short-period cut-of\mb f is not the true minimum period but purely an age ef\mb fect (e.g.\ King \& Schenker 2002). This would arise if systems that we currently observe have not had suf\mb f\mb icient time to evolve to the true period bounce, as illustrated in Figure~\ref{fig:agelim}. Here systems at the currently observed short-period cut-of\mb f of around 77 minutes will continue to evolve to shorter periods for around another $8 \times 10^8$ years before reaching the period bounce (if $\dot{J}_{sys}=\dot{J}_{GR}$, $\lambda = 1.0$). We obtain Pr = 0.659 for the period distribution generated from the single evolutionary track corresponding to Figure~\ref{fig:agelim}, cut at 77 minutes, (for a 1 minute binning, table \ref{tab:tabcomb}, model B), quite close to the value for a f\mb lat distribution. In this period region $\dot{M}\simeq const$ (see Figure \ref{combined}, middle frame).  As $\dot{P}$ scales roughly as $\dot{M}$, the discovery probability is roughly constant if $\gamma=1$.

The same f\mb lat distribution would be obtained if some mechanism would cause systems to 'die' (e.g.\ become too faint to be detected) before reaching the period bounce. Meyer and Meyer-Hofmeister (1999a) speculate that AM Her stars become propeller systems before the period bounce, and so are no longer observed as CVs as their accretion luminosity would be very low. For non-magnetic disc-accreting CVs Meyer and Meyer-Hofmeister (1999b) speculated that as the secondaries become degenerate, the magnetic activity of the secondary reduces rapidly to zero. The disc would then be fed by non-magnetic material, thus reducing the viscosity of the disc plasma and vastly increasing the recurrence time. 

\subsubsection{CAML ef\mb f\mb iciency and primary mass spectrum}
\label{fullcam}
\begin{figure}
  \psfig{file=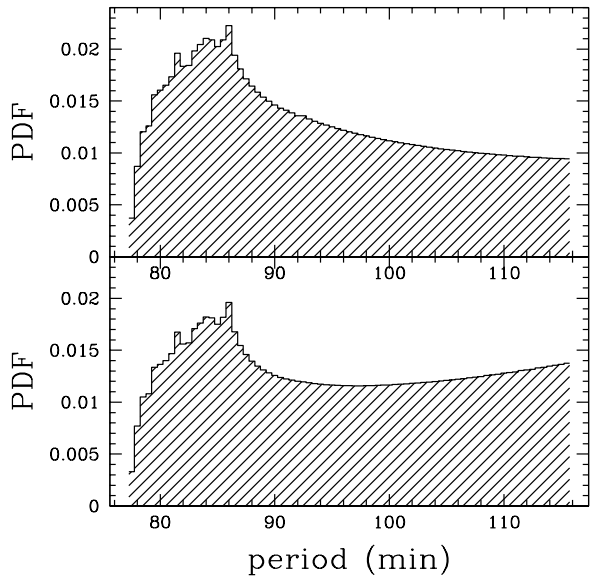,width=\columnwidth}
  \caption{
     Period distribution for a population based on a CAML spectrum ($0\le\eta\le0.95$). Upper frame: $\gamma=1.0$. Lower frame: $\gamma=3.0$. 
    }
  \label{fig:fullcamlgamma}
\end{figure}

We now relax assumption (4) and allow systems to occur with equal probability with any value of the CAML ef\mb f\mb iciency. This produces the period distribution in Figure~\ref{fig:fullcamlgamma} (upper frame) for $\gamma=1$. A spike is still present, though now broadened, and peaked at around 87 minutes. The probability distribution function (PDF) then falls with increasing period. This parent distribution gives Pr=0.831 (1 minute binning, see Table 1, model C1), somewhat larger than that for the f\mb lat distribution.

The result of varying $\gamma$ is summarised in Table \ref{tab:gamma}. At $\gamma=3$ the rejection probability reaches a minimum value Pr = 0.61. This corresponds to the parent distribution shown in Figure~\ref{fig:fullcamlgamma} (lower frame). The broadened peak at around 87 minutes is almost the same as for $\gamma=1$, but at longer $P$ the PDF increases again, i.e.\ there is a minimum at around 95 min. This is caused by a corresponding minimum of $\dot{M}$ along the tracks of Figure \ref{fig:fullcaml}.

\begin{table} 
\caption{$\chi^2$ test on CAML parent distribution versus total observed systems for various $\gamma$} 
\label{tab:gamma}
\begin{center}
\begin{tabular}{ccc}
\hline
 $\gamma$ & $\chi^2_{obs}$ & rejection probability\\
\hline
1 & 1.253 & 0.831\\
2 & 1.167 & 0.751\\
3 & 1.065 & 0.609\\
4 & 1.106 & 0.661\\
6 & 1.152 & 0.957\\
\hline
\end{tabular}
\end{center}
\end{table}

So far we have assumed that all CVs in the population have the same WD mass. Observations (e.g.\ Ritter and Kolb 1998) and population synthesis (e.g.\ de Kool 1992) show that a spread of WD masses is likely. To investigate the ef\mb fect this has on the shape of the PDF near $P_{min}$ we relax assumption 1 and adopt the WD mass spectrum calculated by de Kool (1992).

The corresponding full parent distributions for $\gamma=1$ and $\gamma=3$ give Pr = 0.916 and Pr = 0.691, respectively, for a 1 minute binning (see Table \ref{tab:tabcomb} for full results). These values are slightly higher (worse) than for the CAML ef\mb f\mb iciency spectrum population.

\begin{figure}
  \psfig{file=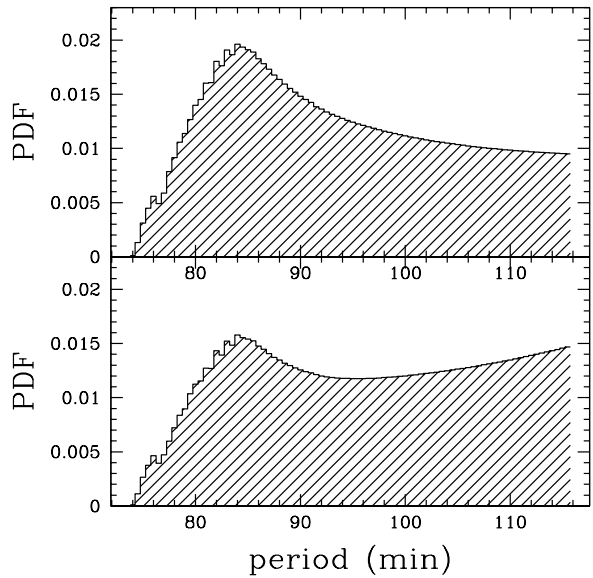,width=\columnwidth}
  \caption{ 
    Period distribution for a population based on a primary mass spectrum (de Kool 1992) and CAML ef\mb f\mb iciency spectrum ($0\le\eta\le0.95$). Upper frame: $\gamma=1$. Lower frame: $\gamma=3$. 
    }
  \label{fig:m1caml}
\end{figure}

If we combine the ef\mb fect of the primary mass distribution and the CAML ef\mb f\mb iciency spectrum, i.e.\ relax assumptions 4 and 1, we obtain the parent distributions shown in Figure \ref{fig:m1caml}. The PDF for $\gamma=1$ (upper frame) gives Pr = 0.786 and exhibits a broad peak with a maximum at around 85 minutes, followed by a gradual decrease with increasing period. The PDF for $\gamma=3$ Figure \ref{fig:m1caml} (lower frame) gives Pr = 0.580 and also shows a similar, though somewhat sharper, broad peak as for the case with constant WD mass. The values of Pr are lower (better) than either of the previous models alone (see Table \ref{tab:tabcomb} for full results). With $\gamma=3$ we approach a value similar to that of a f\mb lat distribution (0.552). 

\subsubsection{Deformation factor spectrum}
\label{bloatf}

\begin{figure}
  \psfig{file=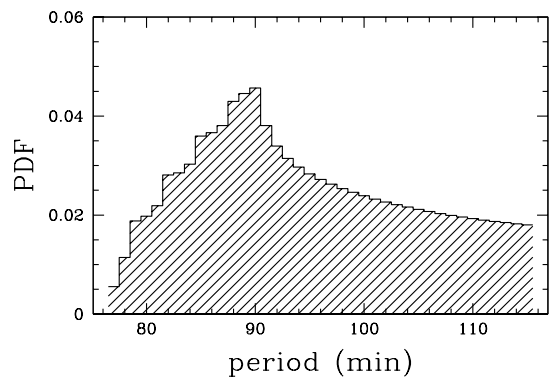,width=\columnwidth}
  \caption{
	Period distribution for a population based on a deformation factor spectrum. 
}
\label{fig:bloating}
\end{figure}

Here we we abandon assumption (5) and assume instead that the secondary stars are subject to various deformation factors $\lambda$ (as described previously in section \ref{baraffe1}). A minimum value of $\lambda=1.18$ is used to set $P_{min}$ equal to the observed $P_{min}=77min$. Any $\lambda$ between 1.18 and a maximum value $\lambda_{max}$ has equal weight. The rejection probability for dif\mb ferent $\lambda_{max}$ are given in Table \ref{tab:bloatingall}. There is a minimum in rejection probability
(Pr=0.887) at $\lambda_{max}\simeq1.35$. The parent distribution generated for this value of $\lambda_{max}$ is shown in Figure \ref{fig:bloating} and exhibits a gradually increasing PDF with increasing period to a peak at around 90 minutes, and then a gradual decrease. 

\begin{table}
\caption{$\chi^2$ test for the model based on a deformation factor spectrum (versus total observed systems, for $\gamma=1.0$)} 
\label{tab:bloatingall}
\begin{center}
\begin{tabular}{ccc}
\hline
maximum & & \\
deformation factor & $\chi^2_{obs}$ & rejection probability\\
\hline
1.31 & 1.392 & 0.934\\
1.33 & 1.347 & 0.915\\
1.35 & 1.297 & 0.887\\
1.37 & 1.382 & 0.928\\
1.39 & 1.452 & 0.955\\
\hline
\end{tabular}
\end{center}
\end{table}

\subsubsection{Initial secondary mass spectrum}
\label{secondary}

\begin{figure}
  \psfig{file=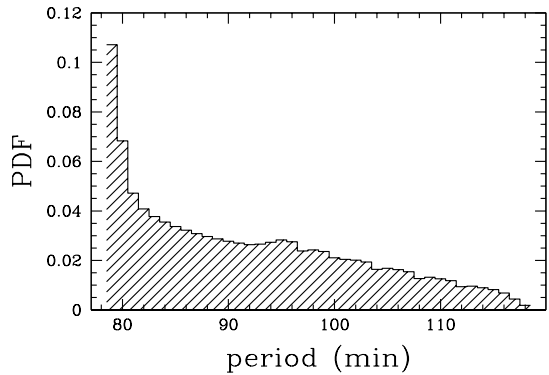,width=\columnwidth}
  \caption{ 
    Period distribution for a population based on an initial secondary mass spectrum. 
    }
  \label{figflagdis}
\end{figure}

We now replace assumption (2) that all systems form with orbital periods greater than 2 hours with the other extreme: all CVs form with orbital periods of less than 2 hours. Specif\mb ically, we assume that all CVs form with donor masses in the range $0.13M_{\odot}\le M_2\le 0.17M_{\odot}$ (this sets $P_{max}=116$ min), and that any $M_2$ is equally likely. From this we obtain a parent distribution as in Figure \ref{figflagdis}. The PDF exhibits a sharp spike at the minimum period (here 78 minutes) and then a gradual decline with increasing period. The corresponding $\chi^2$ test results are given in Table \ref{tab:flagdis}. We conclude that if we were to include any secondary mass spectrum in the previous models we would ef\mb fectively weight the PDF with a ramp function, similar to the one seen on the right of Figure \ref{figflagdis}. This would only increase the rejection probability. 

\begin{table}
\caption{$\chi^2$ test for the model based on a secondary mass spectrum ($0.13M_\odot\le M_2\le0.17M_\odot$) (versus total observed systems, for $\gamma=1$)} 
\label{tab:flagdis}
\begin{center}
\begin{tabular}{ccc}
\hline
binning (min) & $\chi^2_{obs}$ & rejection probability\\
\hline
1 & 3.0878 & $>1-10^{-4}$\\
2 & 4.3233 & $>1-10^{-4}$\\
\hline
\end{tabular}
\end{center}
\end{table}

\subsubsection{A contrived weighting?}

King, Schenker \& Hameury (2002) constructed a (nearly) f\mb lat period distribution by superimposing individual idealized PDFs with dif\mb ferent bounce periods $P_b$ according to a suitably tailored weighting. For the double box-shaped idealised PDFs modelled on the PDF shown in our Figure~\ref{combined} (lower frame) the required weighting is $n(P_b)=\mbox{exp}[-0.124(P_b-P_0)]$ ($P_0$ is the observed minimum period). This weighting function ef\mb fectively mirrors the shape of the sharply peaked individual PDFs. King et al.\ (2002) found that the range $78\le P_b\le93$ is suf\mb f\mb icient to wash out the period spike. It is clear that this procedure involves a certain degree of f\mb ine-tuning for $n(P_b)$ if the shape of the input PDF is given. Such a f\mb ine-tuning must surprise as the two functions involved presumably represent two very dif\mb ferent physical ef\mb fects.

\begin{figure}
  \psfig{file=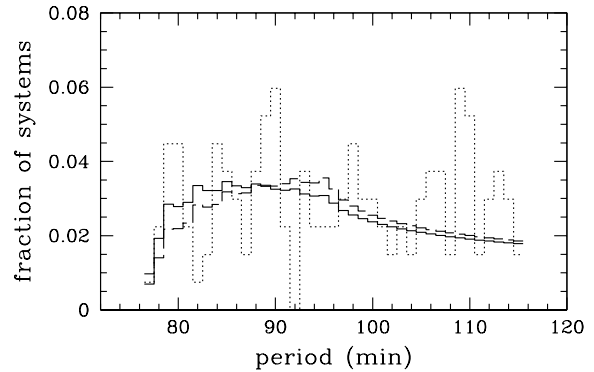,width=\columnwidth}
  \caption{
	Period distributions based on a deformation factor spectrum with $1.18\le\lambda\le1.42$ and $n(P)=\mbox{exp}[-0.124(P_b-P_0)]$ (solid line), $n(P)=\mbox{exp}[-0.07(P_b-P_0)]$ (dashed line). The observed distribution (dotted line) is shown for comparison. 
}
\label{fig:kings}
\end{figure}

We applied the weighting $n(P_b)$ quoted in King et al.\ (2002) to our non-idealized model PDFs that involve the CAML ef\mb f\mb iciency and the deformation factor as a means to vary $P_b$. The weighting produced a marginally worse f\mb it ($P_r=0.841$ versus $P_r=0.831$; 1 minute binning)  for the CAML PDFs compared to the parent population based on a f\mb lat CAML ef\mb f\mb iciency spectrum we discussed earlier. In part this is due to the fact that the upper limit on $\eta$ does not allow a big enough range of $P_b$. In the case of the deformation factor PDFs the f\mb it marginally improved ($P_r=0.837$ versus $P_r=0.887$; 1 minute binning, $1.18\le\lambda\le1.35$). It is possible to optimise the f\mb it by adding systems with deformation factors up to 1.42, and by using the weighting $n(P_b)=\mbox{exp}[-0.07(P_b-P_0)]$, but this still gives the fairly large value  $P_r = 0.829$ (see also Figure \ref{fig:kings}). However, such a parent population is inconsistent with the observed distribution for longer periods. As can be seen from Figure \ref{fig:barraffe} systems that are subject to larger deformation factors would evolve into the period gap, hence the gap would be populated in this model. 

\begin{figure}
  \psfig{file=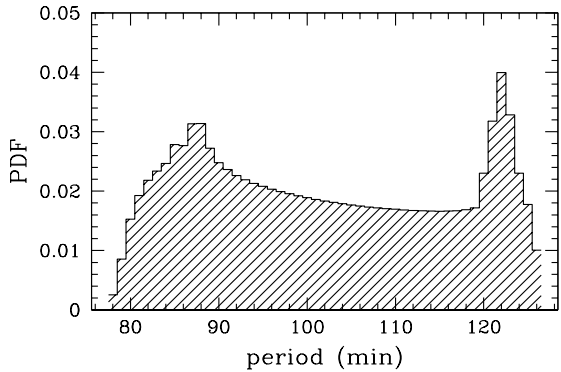,width=\columnwidth}
  \caption{
	Period distributions based on a spread of systemic AM losses $5-11\times\dot{J}_{GR}$ and $n(P)=exp[-0.124(P_b-P_O]$ 
}
\label{fig:grsum1}
\end{figure}

For completeness we show in Figure \ref{fig:grsum1} the result of the superposition suggested by King, Schenker \& Hameury (2002) if realistic rather than idealised PDFs are used. This model assumes additional systemic AM losses ($5-11 \times \dot J_{GR}$; no CAML, no deformation factor, $\gamma =1$) as the control parameter for varying $P_b$, and the weighting as in King et al. The pronounced feature just above 2 hrs orbital period is the result of the adiabatic reaction of the donor stars at turn-on of mass transfer (see e.g.\ Ritter \& Kolb 1992). Such a feature is absent in the observed distribution. If deformation ef\mb fects are taken into account the additional AM losses required to wash out the $P_{min}$ spike would cover a similar range but at a smaller magnitude. The resulting period distribution would be similar to the one shown in Figure \ref{fig:grsum1}

\section{discussion}
\label{discuss}

We have investigated mechanisms that could increase the bounce period for CVs from the canonical theoretical value $\sim65$ min to the observed value $P_{min}\approx77$ min, and ways to wash out the theoretically predicted accumulation of systems near the minimum period (the period spike). Unlike King, Schenker \& Hameury 2002 we focussed on ef\mb fects other than increased systemic angular momentum (AM) losses, i.e. we assume that gravitational radiation is the only systemic sink of orbital AM.

We f\mb ind that even a maximal ef\mb f\mb icient consequential AM loss (CAML) mechanism cannot increase the bounce period suf\mb f\mb iciently. As the real CV population is likely to comprise systems with a range of CAML ef\mb f\mb iciencies we would in any case expect to have a distribution of systems down to $\sim65$ min, rather than the observed sharp cut-of\mb f.

We considered donor stars that are ``bloated'' due to intrinsic ef\mb fects, such as the tidal deformation found in 3-dim.\ SPH simulations of Roche-lobe f\mb illing stars. An implausibly large deformation factor of around 1.18 is needed to obtain a bounce period of $\sim77$~min. 

A possible alternative identif\mb ication of $P_{min}$ as an age limit rather than a period bounce (King \& Schenker 2002) would limit the donor mass in any CV in a CV population dominated by hydrogen--rich, unevolved systems to $>0.1M_{\odot}$. Any system with donor mass much less than this would either have an orbital period less than 78 minutes or would have already evolved beyond the period minimum. There are indeed systems with suspected $M_2<0.1M_{\odot}$; good candidates are WZ Sge ($M_2\simeq0.058$; Patterson et al 1998) and OY Car ($M_2\simeq0.07M_{\odot}$; Pratt et al.\ 1999). 

It is also possible that systems 'die' or fade before reaching the period bounce, and hence become undetectable as CVs. The fact that the very dif\mb ferent groups of non-magnetic and magnetic CVs show almost identical values of $P_{min}$ (see Figure \ref{fig:mnmcvs}) strongly suggests that the physical cause for the potential fading would have to be rooted in the donor stars or the evolution rather than the accretion physics or emission properties of the systems. 
 
Even if the bounce period problem is ignored we f\mb ind in all synthesized model populations (except for the age limit model) a pronounced remaining feature due to the accumulation of systems near the bounce. We employ a modif\mb ied $\chi^2$ test to measure the ``goodness'' of f\mb it against the observed sample. An F-test (Press et al 1992) was also applied to the majority of $\gamma=1$ models and the same general trends observed. None of our synthesised model populations f\mb its as well as the distribution which is simply f\mb lat in orbital period (rejection probability $P_r \simeq 55\%$). Only models where brighter systems carry a far greater weight than expected in a simple magnitude--limited sample (selection factor $\propto \dot M^\gamma$ with $\gamma \simeq 3$ rather than $\simeq 1$) achieve similar values for $P_r$. However, most of our models with $\gamma = 1$ canot be rejected unambiguously on the basis of this test. 

Models designed to ``wash out'' the period spike by introducing a large spread of the CAML ef\mb f\mb iciency do generally better than population models based on donor stars that are subject to a large spread of intrinsic deformation factors. For all models the rejection probability decreases if the full WD mass spectrum is taken into account, as this introduces an additional spread in the bounce period. Model populations where all CVs form at long orbital periods (chief\mb ly above the period gap) give a much better f\mb it than models that include newborn CVs with small donor mass. Adding these systems to the population introduces a general increase of the orbital period distribution towards short periods, thus making the period spike more pronounced. This suggests that most CVs must have formed at long periods and evolved through the period gap to become short-period CVs. This is consistent with independent evidence that CV secondary stars are somewhat evolved (Baraf\mb fe \& Kolb 2000; Schenker et al.\ 2002; Thorstensen et al 2002). 

Recently, King, Schenker \& Hameury (2002) constructed a f\mb lat orbital period distribution by superimposing idealised PDFs that describe subpopulations of CVs with a f\mb ixed initial donor mass and initial WD mass, but dif\mb ferent bounce periods. This superposition required a strongly declining number of systems with increasing bounce periods. We repeated this experiment with a realistic PDF, but failed to obtain a markedly improved f\mb it.  

In conclusion, we f\mb ind that the period minimum problem and the period spike problem remain an open issue. It is possible to construct CV model populations where the period spike is washed out suf\mb f\mb iciently so that it cannot be ruled out unambiguously on the basis of an objective statisticial test against the observed CV period distribution.

\section{Acknowledgements}

We thank Graham Wynn, Andrew King and Isabelle Baraf\mb fe for useful discussions. Andrew Conway and Chris Jones who gave advice on the statistical analysis. We also thank Andrew Norton for a critical reading of the paper and the referee
 Jean-Marie Hameury for useful comments.

\end{document}